\documentclass[aip,apl,reprint,twocolumn,a4paper,showkeys,balancelastpage]{revtex4-1}

\usepackage[version=3]{mhchem} 
\usepackage{color}
\usepackage{xr}

\newcommand{\unit}{\ \mathrm}
\newcommand{\temp}[1]{$#1~^{\circ}$C}

\graphicspath{{./images/}}

\begin{document}

\title{Ballistic transport in graphene grown by chemical vapor deposition}

\author{V. E. Calado}
\altaffiliation{Contributed equally to this work}
\affiliation{Kavli Institute of Nanoscience, Delft University of Technology, 2600 GA Delft, The Netherlands}

\author{Shou-En~Zhu}
\altaffiliation{Contributed equally to this work}
\affiliation{Micro and Nano Engineering Laboratory, Precision and Microsystems Engineering, Delft University of Technology, 2628 CD Delft, The Netherlands}

\author{S.~Goswami}
\affiliation{Kavli Institute of Nanoscience, Delft University of Technology, 2600 GA Delft, The Netherlands}

\author{Q.~Xu}
\affiliation{Kavli Institute of Nanoscience, Delft University of Technology, 2600 GA Delft, The Netherlands}

\author{K.~Watanabe}
\affiliation{Advanced Materials Laboratory, National Institute for Materials Science, 1-1 Namiki, Tsukuba, 305-0044, Japan}

\author{T.~Taniguchi}
\affiliation{Advanced Materials Laboratory, National Institute for Materials Science, 1-1 Namiki, Tsukuba, 305-0044, Japan}

\author{G.~C.~A.~M.~Janssen}
\affiliation{Micro and Nano Engineering Laboratory, Precision and Microsystems Engineering, Delft University of Technology, 2628 CD Delft, The Netherlands}

\author{L.~M.~K.~Vandersypen}
\email{l.m.k.vandersypen@tudelft.nl}
\affiliation{Kavli Institute of Nanoscience, Delft University of Technology, 2600 GA Delft, The Netherlands}

\begin{abstract}

In this letter we report the observation of ballistic transport on micron length scales in graphene synthesised by chemical vapour deposition (CVD). Transport measurements were done on Hall bar geometries in a liquid He cryostat. Using non-local measurements we show that electrons can be ballistically directed by a magnetic field (transverse magnetic focussing) over length scales of $\sim1\unit{\mu m}$.  Comparison with atomic force microscope measurements suggests a correlation between the absence of wrinkles and the presence of ballistic transport in CVD graphene.
\end{abstract}

\maketitle

\setcounter{table}{0}
\renewcommand{\thetable}{\arabic{table}}%
\setcounter{figure}{0}
\renewcommand{\thefigure}{\arabic{figure}}%

High electronic quality in graphene is a key requirement for many experiments and future applications \cite{Novoselov2012}. The highest quality has so far been achieved in exfoliated graphene\cite{Novoselov2004}, either by suspending the graphene flakes \cite{bolotin_ultrahigh_2008} or by depositing them on hexagonal boron nitride (hBN) substrates\cite{Dean2010}. To move beyond a laboratory setting, mass production of graphene is essential. Among several promising synthesis methods, chemical vapor deposition (CVD) is a low-cost, scalable and controllable method for the production of monolayer graphene \cite{Mattevi2011,Zhang2013}. Using CVD, large and predominantly monolayer graphene of high quality has been synthesized on copper foils\cite{Li2009c}. Considerable effort has been made to scale up the technology to produce meter-sized foils \cite{Bae2010,Kobayashi2013} and to achieve crystals with $\sim$mm dimensions \cite{Li2011,Yan2012,Wu2012}. Such large crystal sizes minimize short-range scattering from grain boundaries \cite{Yu2011,Koepke2012,Li2010,Yazyev2010}. Also for CVD graphene, the highest electronic quality is realized by transferring it onto hBN \cite{Petrone2012,Gannett2011}, using a clean (contaminant-free) and dry procedure \cite{Zomer2011,Dean2010}.

Despite this effort the electronic quality of CVD graphene is still considered to be inferior to that of exfoliated graphene, and, in particular, there are no reports of ballistic transport phenomena in CVD graphene. Ballistic transport is of relevance for realizing electron optics experiments such as Veselago lensing\cite{Cheianov2007} and angle-resolved Klein tunnelling\cite{Katsnelson2006} or specular Andreev reflection\cite{Beenakker2006}.
A negative bend resistance in a cross geometry \cite{Hirayama1991,Mayorov2011} gives a first indication of ballistic transport. A more sensitive probe, since it is more easily affected by small angle scattering, is transverse magnetic focussing (TMF), seen two decades ago in a GaAs/AlGaAs 2-dimensional electron gas\cite{Houten1989} and only recently in exfoliated graphene\cite{Taychatanapat2013}.

\begin{figure}
    \centering
    \includegraphics{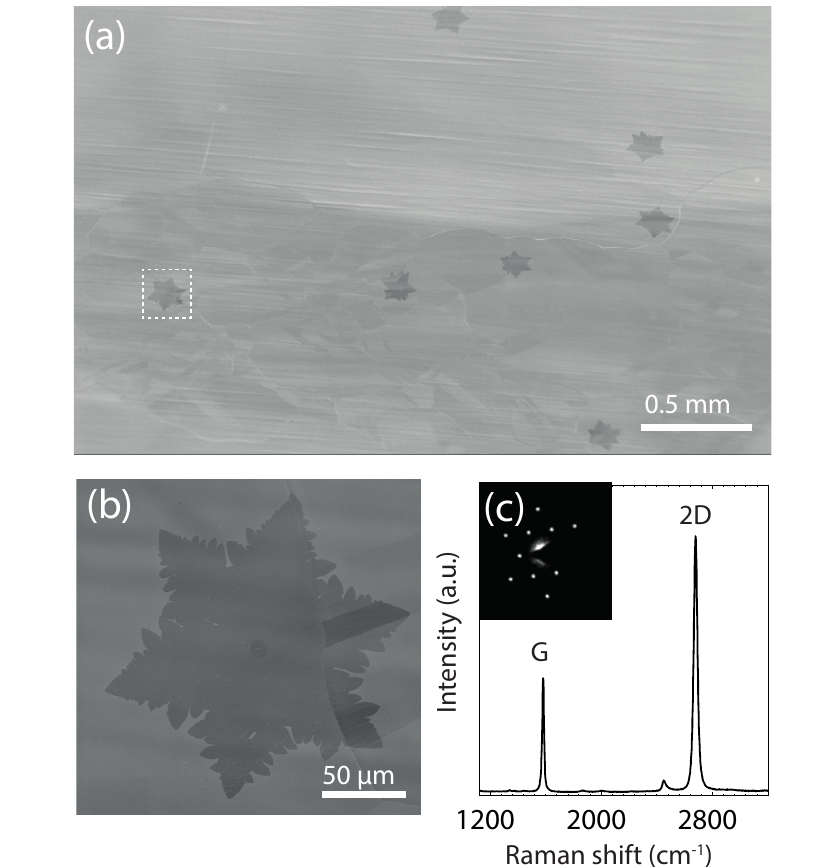}%
    \caption{(a) Scanning electron microscope (SEM) image of a copper foil with isolated graphene crystals after a short growth time. (b) An SEM image of one of the crystals in (a), where the dendritic shape at the edges is visible. The dark stripes are different crystal orientations in the Cu. (c) Raman spectrum taken of another graphene flake transferred from copper to SiO$_{2}$, grown in similar conditions. Inset: a typical diffraction pattern, recorded in a transmission electron microscope (TEM). \label{BallisticCVD:fig1}}%
\end{figure}

Here we report ballistic transport in graphene grown by the CVD method.  Large grain size single crystals are grown on a folded copper foil enclosure and subsequently dry-transferred onto hBN flakes. Ballistic transport is demonstrated in the form of transverse magnetic focussing (TMF).

Following the work of Refs. \cite{Li2009c,Li2011}, a copper foil with a thickness of $25\unit{\mu m}$ is cut in $\sim2\times3\unit{cm^2}$ sheets  (Alfa Aesar $>99.8\%$ pure). The foil is folded to form a fully enclosed pocket and placed inside a quartz tube in a home built tube oven. $0.5\unit{sccm}$ $\rm{CH_4}$ and $2\unit{sccm}$ $\rm{H_2}$ is fed through the tube with a $\rm{CH_4}$ partial pressure of less than $20\unit{\mu bar}$. The temperature is set to $1050\temp$, close to the Cu melting point. With these parameters we obtain a low nucleation density in the inside of the foil pocket. In Fig.~\ref{BallisticCVD:fig1}a we show a scanning electron microscope (SEM) image of seven graphene crystals on copper spread over an area of $3.1\times2.0\unit{mm^2}$. This yields a nucleation density of $\sim1.1\unit{mm^{-2}}$. With such a low nucleation density we are able to grow crystals that have an average diameter of $\sim1\unit{mm}$. However, in this letter we used isolated crystals, formed in the early stage of CVD growth, such as the one shown in Fig.~\ref{BallisticCVD:fig1}b. These crystals are about $150\unit{\mu m}$ across and are grown in about 30 min. The crystals have a sixfold dendritic shape. We note that at the nucleation site, a second layer starts growing.

In Fig.~\ref{BallisticCVD:fig1}c we show a Raman spectrum taken on a graphene crystal similar to those in Fig.~\ref{BallisticCVD:fig1}a, after transfer to SiO$_{2}$. The spectrum confirms that the crystals are monolayer graphene\cite{Ferrari2006} with a defect density below the Raman detection limit, as no D line at $\sim1350\unit{cm^{-1}}$ is visible. In the inset of Fig.~\ref{BallisticCVD:fig1}c a transmission electron microscope (TEM) diffraction pattern is shown. It confirms a hexagonal lattice~\cite{Meyer2007}. We have recorded many more diffraction patterns\cite{Supplement}, which show the same lattice orientation over a distance of $\sim50\unit{\mu m}$. This indicates that the graphene patches in Fig.~\ref{BallisticCVD:fig1}a and Fig.~\ref{BallisticCVD:fig1}b are monocrystalline, i.e. have no grain boundaries. CVD graphene on copper is transferred onto a hBN flake\cite{Supplement}. The hBN flakes are prepared by mechanical exfoliation on a polymer substrate. A 250 nm thick hBN is selected and transferred by the method described in Ref.\cite{Zomer2011} onto e-beam defined tungsten (W) gate electrodes, so that the hBN acts as a gate dielectric. The tungsten can withstand high temperatures ($\sim600\temp$) during annealing to remove residues from the hBN flake. Furthermore the bottom gate screens charged impurities, presumably present in the SiO$_2$ below.

\begin{figure}
 \centering
 \includegraphics{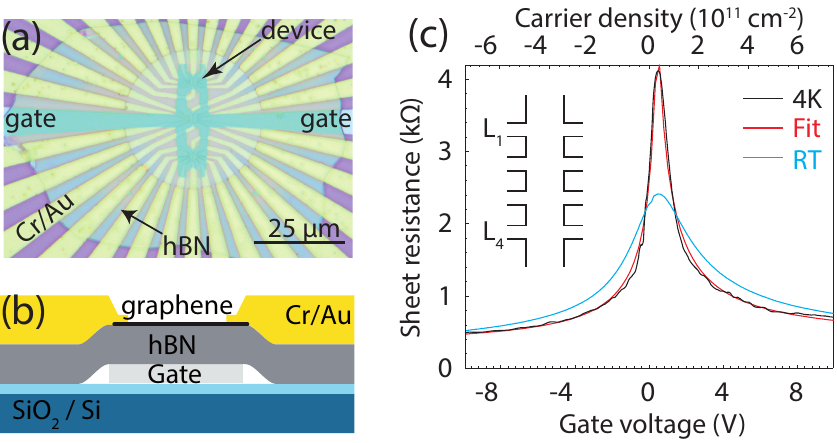}%
 \caption{(a) Optical microscope image of the device. A hBN flake (light blue) is transferred onto a tungsten bottom gate. Yellow stripes are the gold contacts. They are deposited in two steps, with the outer part much thicker ($\sim 350$ nm) than the inner part, leading to the circular pattern in the image. (b) A schematic side view of the device, with the materials indicated. The hBN acts here as a dielectric between the graphene and W bottom gate. (c) Black and cyan: the sheet resistance as a function of gate voltage and carrier density taken at 4~K respectively room temperature between probes L$_1$ and L$_4$ (see inset). Red: a fit to the 4K data using the self-consistent  equation for diffusive transport as a model.\label{BallisticCVD:fig2}}%
\end{figure}

We have contacted the CVD graphene flake with e-beam lithography defined $3\unit{nm}$ Cr / $25\unit{nm}$ Au contacts (Fig.~\ref{BallisticCVD:fig2}a) and subsequently etched Hall bars with reactive ion etching in oxygen. In Fig.~\ref{BallisticCVD:fig2}b we give a device schematic.

Transport measurements were done in vacuum at 4~K and at room temperature (RT). In Fig.~\ref{BallisticCVD:fig2}c we show the sheet resistance measured at 4~K in black and at RT in cyan. The resistance peak at the charge neutrality point (CNP) became taller and narrower upon cooling as expected. We applied a $1\unit{\mu A}$ dc current bias across the Hall bar (in the inset) and measured the voltage between terminals $\rm{L_1}$ and $\rm{L_4}$  as a function of gate voltage on the tungsten bottom gate. The charge carrier density is tuned by the gate voltage with a coupling strength of $7.45\pm0.02\cdot10^{10}\unit{cm^{-2}~V^{-1}}$, extracted from Hall measurements\cite{Supplement}. We find the CNP is offset by only $3.31\cdot10^{10}\unit{cm^{-2}}$, indicating very little background doping.

We characterize the transport properties of the device by fitting the 4K data with the self-consistent Boltzmann equation for diffusive transport that includes long and short range scattering\cite{Adam2007,Hwang2007}: $\rho=\left( ne\mu_c +\sigma_0\right)^{-1}+\rho_S$, where $\mu_c$ is the mobility from long range scattering, $\sigma_0$ the minimum conductivity at the CNP and $\rho_s$ the resistivity from short range scattering. This model fits very well to the data when we account for the electron-hole asymmetry by using different fit parameters for the two sides. For the low temperature hole mobility we find $\mu_h=41500\pm800\unit{cm^2~V^{-1}~s^{-1}}$, for the electron mobility $\mu_e=28700\pm600\unit{cm^2~V^{-1}~s^{-1}}$. The RT mobilities are about a factor two lower \cite{Supplement}.

For the resistivity from short range scattering we obtain $\rho_S=280\pm10\unit{\Omega}$ for holes and $\rho_S=380\pm10\unit{\Omega}$ for electrons. These are higher than what has been found earlier for exfoliated flakes on hBN ($\sim70\unit{\Omega}$) \cite{Dean2010}. For the residual conductivity $\sigma_0$ we find a value of $221\pm1.5\unit{\mu S}$, which is $5.70 \pm 0.04~e^2/h$. The mobility values found here are for long range scattering only. Using the Drude model of conductivity one finds a mean free path of $200-400\unit{nm}$ for a density of $7\cdot10^{11}\unit{cm^{-2}}$.

\begin{figure}
   \centering
 \includegraphics{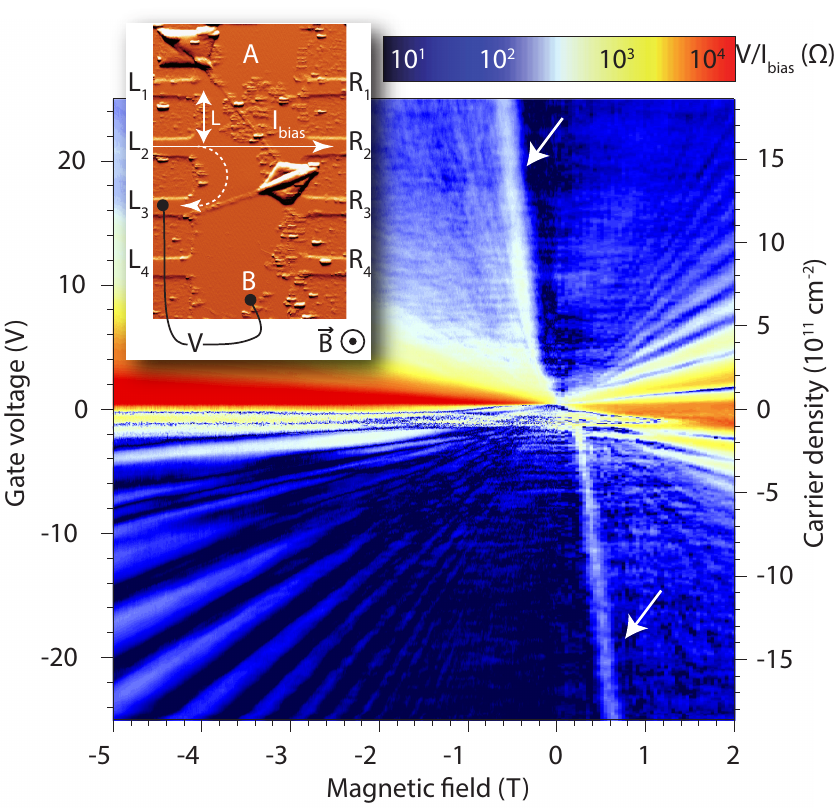}%
 \caption{The resistance $V/I_{bias}$ as function of gate voltage and magnetic field plotted in a logarithmic color scale. The straight lines are due to SdH oscillations and the square-root line is due to TMF, indicated by arrows. Inset: AFM image of the measured device, showing the non-local measurement configuration used for observing magnetic focussing. $V$ is measured between $\rm{L_3}$ and $B$ while a current bias is applied between L$_2$ and R$_2$. \label{BallisticCVD:fig3}}%
\end{figure}

Next, we test whether this device allows transverse magnetic focusing (TMF). The observation of TMF would directly imply the occurrence of ballistic transport in that part of the device. As shown in the inset of Fig.~\ref{BallisticCVD:fig3}, we apply a magnetic field perpendicular to the device with a current bias from contact $\rm{L_2}$ to $\rm{R_2}$. The Lorentz force will act on the charge carriers and will steer them in a circular orbit with cyclotron radius $R_c=\hbar k_F/eB$, where $R_c$ is the cyclotron radius. Electrons leaving contact $\rm{L_2}$ can reach contact $\rm{L_3}$ when the cyclotron radius matches one half the distance between the contacts $L$, provided the electrons are not scattered while traveling along the semi-circle joining the contacts. This focussing condition occurs for specific combinations of magnetic field and gate voltage:
\begin{equation}
\label{eq:focus}
B=\frac{2\hbar k_F}{eL}\propto\sqrt{V_{gate}}.
\end{equation}
The momentum of the charges, $\hbar k_F$, is tuned with the bottom gate voltage $V_{gate}$.

When electrons reach contact $\rm{L_3}$, its potential will be raised. We probe this potential by recording the voltage $V$ between terminal L$_3$ and B, making the assumption that the potential of the far-away contact B remains constant. The gate voltage and magnetic field are swept and the resistance $V/I_{bias}$ is plotted on a logarithmic color-scale in Fig.~\ref{BallisticCVD:fig3}.

In Fig.~\ref{BallisticCVD:fig3}, above fields of $\pm$1 T Shubnikov-de Haas (SdH) oscillations are seen as straight lines diverging for larger magnetic field. However the lines marked with white arrows do not fit the SdH pattern. These lines are attributed to TMF, following from the square-root dependence between the magnetic field and charge carrier density, see Eq.~1. For negative field and positive gate voltage, electrons leaving $\rm{L_2}$ are deflected towards contact $\rm{L_3}$, where an increase of the voltage $V$ is observed. For positive field and negative gate voltage, we find the same square-root dependence, where holes are deflected instead of electrons. Such behavior can only be observed when the region between the contacts permits ballistic transport, i.e. little scattering takes place. The fact that the focussing signal appears only for a specific combination of magnetic field and gate voltage, giving rise to a sharp peak in the plot, illustrates the sensitivity of the focussing effect to scattering: only small deviations away from the proper semi-circular path are enough to make electrons miss contact $\rm{L_3}$.

From a fit to the data, we find a distance of $L=570\unit{nm}$, which corresponds to a semicircle distance of $\sim900\unit{nm}$ that electrons have travelled. This value is somewhat larger than the expected lithographic distance \cite{Supplement} of $500\unit{nm}$. A similar mismatch between the extracted contact separation from TMF and the lithographic distance was recently reported in exfoliated graphene \cite{Taychatanapat2013}. In these experiments, TMF peaks were seen also at integer multiples of the B value given in Eq.~1. The absence of such peaks in our measurements suggests that the edges in our sample are too rough or dirty to allow specular reflection.

We have performed analogous measurements across the entire device and except between $\rm{L_1}$ and $\rm{L_2}$ no focussing was found in other parts \cite{Supplement}, pointing at the presence of inhomogeneities.
In the inset of Fig.~\ref{BallisticCVD:fig3} we show an atomic force microscope (AFM) image of the device. In the device a bubble and two wrinkles are present, which may hinder transport. The wrinkles are $1-2\unit{nm}$ in height and less than $\sim40\unit{nm}$ in width (not accounting for convolution with the AFM tip width). The bubbles are about $\sim35\unit{nm}$ high. The apparent connection between the absence of wrinkles or bubbles and the observation of TMF in this device indicates that is worth investigating more systematically whether such wrinkles and bubbles are sufficient to spoil ballistic effects. More measurements and further discussion of the inhomogeneity of the device is found in Ref. \cite{Supplement}. We note that given the presence of the wrinkles and bubbles in the Hall bar, the mobility values are remarkably high, in the range of the highest reported mobilities in CVD graphene\cite{Petrone2012}.

In summary, we have demonstrated that TMF can be observed in CVD graphene on a micron scale distance. The CVD process was optimized to obtain large single crystal flakes. In order to preserve its high quality we have transferred graphene flakes with a dry method onto hBN. The main limitation in electronic quality for the current device appears to be the presence of wrinkles and bubbles. If one optimizes further the processing to reduce or eliminate the wrinkles and bubbles, it may become possible to routinely observe ballistic phenomena in CVD graphene. Already, the present results are an import step forward in the direction of scalable and controllable graphene production not only for industrial applications but also for fundamental research involving ballistic effects in graphene.
\\

We thank A.M.~Goossens for discussions and R.~Schouten, R.~Luttjeboer, P.~van Holst, H.~Jansen, L.~Schipperheijn for technical assistance. We thank G.F.~Schneider and S.R.K.~Malladi for preparing the TEM sample and H.~Zandbergen for providing access to the TEM setup. We acknowledge the Young Wild Idea Grant from the Delft Centre for Materials (DCMat) for financial support. This work is part of the research program of the Foundation for Fundamental Research on Matter (FOM), which is part of the Netherlands Organization for Scientific Research (NWO).


\begin{thebibliography}{31}%
\makeatletter
\providecommand \@ifxundefined [1]{%
 \@ifx{#1\undefined}
}%
\providecommand \@ifnum [1]{%
 \ifnum #1\expandafter \@firstoftwo
 \else \expandafter \@secondoftwo
 \fi
}%
\providecommand \@ifx [1]{%
 \ifx #1\expandafter \@firstoftwo
 \else \expandafter \@secondoftwo
 \fi
}%
\providecommand \natexlab [1]{#1}%
\providecommand \enquote  [1]{``#1''}%
\providecommand \bibnamefont  [1]{#1}%
\providecommand \bibfnamefont [1]{#1}%
\providecommand \citenamefont [1]{#1}%
\providecommand \href@noop [0]{\@secondoftwo}%
\providecommand \href [0]{\begingroup \@sanitize@url \@href}%
\providecommand \@href[1]{\@@startlink{#1}\@@href}%
\providecommand \@@href[1]{\endgroup#1\@@endlink}%
\providecommand \@sanitize@url [0]{\catcode `\\12\catcode `\$12\catcode
  `\&12\catcode `\#12\catcode `\^12\catcode `\_12\catcode `\%12\relax}%
\providecommand \@@startlink[1]{}%
\providecommand \@@endlink[0]{}%
\providecommand \url  [0]{\begingroup\@sanitize@url \@url }%
\providecommand \@url [1]{\endgroup\@href {#1}{\urlprefix }}%
\providecommand \urlprefix  [0]{URL }%
\providecommand \Eprint [0]{\href }%
\providecommand \doibase [0]{http://dx.doi.org/}%
\providecommand \selectlanguage [0]{\@gobble}%
\providecommand \bibinfo  [0]{\@secondoftwo}%
\providecommand \bibfield  [0]{\@secondoftwo}%
\providecommand \translation [1]{[#1]}%
\providecommand \BibitemOpen [0]{}%
\providecommand \bibitemStop [0]{}%
\providecommand \bibitemNoStop [0]{.\EOS\space}%
\providecommand \EOS [0]{\spacefactor3000\relax}%
\providecommand \BibitemShut  [1]{\csname bibitem#1\endcsname}%
\let\auto@bib@innerbib\@empty
\bibitem [{\citenamefont {Novoselov}\ \emph {et~al.}(2012)\citenamefont
  {Novoselov}, \citenamefont {Falko}, \citenamefont {Colombo}, \citenamefont
  {Gellert}, \citenamefont {Schwab},\ and\ \citenamefont
  {Kim}}]{Novoselov2012}%
  \BibitemOpen
  \bibfield  {author} {\bibinfo {author} {\bibfnamefont {K.~S.}\ \bibnamefont
  {Novoselov}}, \bibinfo {author} {\bibfnamefont {V.~I.}\ \bibnamefont
  {Falko}}, \bibinfo {author} {\bibfnamefont {L.}~\bibnamefont {Colombo}},
  \bibinfo {author} {\bibfnamefont {P.~R.}\ \bibnamefont {Gellert}}, \bibinfo
  {author} {\bibfnamefont {M.~G.}\ \bibnamefont {Schwab}}, \ and\ \bibinfo
  {author} {\bibfnamefont {K.}~\bibnamefont {Kim}},\ }\href {\doibase
  10.1038/nature11458} {\bibfield  {journal} {\bibinfo  {journal} {Nature
  (London)}\ }\textbf {\bibinfo {volume} {490}},\ \bibinfo {pages} {192}
  (\bibinfo {year} {2012})}\BibitemShut {NoStop}%
\bibitem [{\citenamefont {Novoselov}\ \emph {et~al.}(2004)\citenamefont
  {Novoselov}, \citenamefont {Geim}, \citenamefont {Morozov}, \citenamefont
  {Jiang}, \citenamefont {Zhang}, \citenamefont {Dubonos}, \citenamefont
  {Grigorieva},\ and\ \citenamefont {Firsov}}]{Novoselov2004}%
  \BibitemOpen
  \bibfield  {author} {\bibinfo {author} {\bibfnamefont {K.~S.}\ \bibnamefont
  {Novoselov}}, \bibinfo {author} {\bibfnamefont {A.~K.}\ \bibnamefont {Geim}},
  \bibinfo {author} {\bibfnamefont {S.~V.}\ \bibnamefont {Morozov}}, \bibinfo
  {author} {\bibfnamefont {D.}~\bibnamefont {Jiang}}, \bibinfo {author}
  {\bibfnamefont {Y.}~\bibnamefont {Zhang}}, \bibinfo {author} {\bibfnamefont
  {S.~V.}\ \bibnamefont {Dubonos}}, \bibinfo {author} {\bibfnamefont {I.~V.}\
  \bibnamefont {Grigorieva}}, \ and\ \bibinfo {author} {\bibfnamefont {A.~A.}\
  \bibnamefont {Firsov}},\ }\href {\doibase 10.1126/science.1102896} {\bibfield
   {journal} {\bibinfo  {journal} {Science}\ }\textbf {\bibinfo {volume}
  {306}},\ \bibinfo {pages} {666} (\bibinfo {year} {2004})}\BibitemShut
  {NoStop}%
\bibitem [{\citenamefont {Bolotin}\ \emph {et~al.}(2008)\citenamefont
  {Bolotin}, \citenamefont {Sikes}, \citenamefont {Jiang}, \citenamefont
  {Klima}, \citenamefont {Fudenberg}, \citenamefont {Hone}, \citenamefont
  {Kim},\ and\ \citenamefont {Stormer}}]{bolotin_ultrahigh_2008}%
  \BibitemOpen
  \bibfield  {author} {\bibinfo {author} {\bibfnamefont {K.}~\bibnamefont
  {Bolotin}}, \bibinfo {author} {\bibfnamefont {K.}~\bibnamefont {Sikes}},
  \bibinfo {author} {\bibfnamefont {Z.}~\bibnamefont {Jiang}}, \bibinfo
  {author} {\bibfnamefont {M.}~\bibnamefont {Klima}}, \bibinfo {author}
  {\bibfnamefont {G.}~\bibnamefont {Fudenberg}}, \bibinfo {author}
  {\bibfnamefont {J.}~\bibnamefont {Hone}}, \bibinfo {author} {\bibfnamefont
  {P.}~\bibnamefont {Kim}}, \ and\ \bibinfo {author} {\bibfnamefont
  {H.}~\bibnamefont {Stormer}},\ }\href {\doibase 10.1016/j.ssc.2008.02.024}
  {\bibfield  {journal} {\bibinfo  {journal} {Solid State Commun.}\ }\textbf
  {\bibinfo {volume} {146}},\ \bibinfo {pages} {351} (\bibinfo {year}
  {2008})}\BibitemShut {NoStop}%
\bibitem [{\citenamefont {Dean}\ \emph {et~al.}(2010)\citenamefont {Dean},
  \citenamefont {Young}, \citenamefont {Meric}, \citenamefont {Lee},
  \citenamefont {Wang}, \citenamefont {Sorgenfrei}, \citenamefont {Watanabe},
  \citenamefont {Taniguchi}, \citenamefont {Kim}, \citenamefont {Shepard},\
  and\ \citenamefont {Hone}}]{Dean2010}%
  \BibitemOpen
  \bibfield  {author} {\bibinfo {author} {\bibfnamefont {C.}~\bibnamefont
  {Dean}}, \bibinfo {author} {\bibfnamefont {A.}~\bibnamefont {Young}},
  \bibinfo {author} {\bibfnamefont {I.}~\bibnamefont {Meric}}, \bibinfo
  {author} {\bibfnamefont {C.}~\bibnamefont {Lee}}, \bibinfo {author}
  {\bibfnamefont {L.}~\bibnamefont {Wang}}, \bibinfo {author} {\bibfnamefont
  {S.}~\bibnamefont {Sorgenfrei}}, \bibinfo {author} {\bibfnamefont
  {K.}~\bibnamefont {Watanabe}}, \bibinfo {author} {\bibfnamefont
  {T.}~\bibnamefont {Taniguchi}}, \bibinfo {author} {\bibfnamefont
  {P.}~\bibnamefont {Kim}}, \bibinfo {author} {\bibfnamefont {K.}~\bibnamefont
  {Shepard}}, \ and\ \bibinfo {author} {\bibfnamefont {J.}~\bibnamefont
  {Hone}},\ }\href {\doibase 10.1038/nnano.2010.172} {\bibfield  {journal}
  {\bibinfo  {journal} {Nat. Nanotechnol.}\ }\textbf {\bibinfo {volume} {5}},\
  \bibinfo {pages} {722} (\bibinfo {year} {2010})}\BibitemShut {NoStop}%
\bibitem [{\citenamefont {Mattevi}, \citenamefont {Kim},\ and\ \citenamefont
  {Chhowalla}(2011)}]{Mattevi2011}%
  \BibitemOpen
  \bibfield  {author} {\bibinfo {author} {\bibfnamefont {C.}~\bibnamefont
  {Mattevi}}, \bibinfo {author} {\bibfnamefont {H.}~\bibnamefont {Kim}}, \ and\
  \bibinfo {author} {\bibfnamefont {M.}~\bibnamefont {Chhowalla}},\ }\href
  {http://dx.doi.org/10.1039/C0JM02126A} {\bibfield  {journal} {\bibinfo
  {journal} {J. Mater. Chem.}\ }\textbf {\bibinfo {volume} {21}},\ \bibinfo
  {pages} {3324} (\bibinfo {year} {2011})}\BibitemShut {NoStop}%
\bibitem [{\citenamefont {Zhang}, \citenamefont {Zhang},\ and\ \citenamefont
  {Zhou}(2013)}]{Zhang2013}%
  \BibitemOpen
  \bibfield  {author} {\bibinfo {author} {\bibfnamefont {Y.}~\bibnamefont
  {Zhang}}, \bibinfo {author} {\bibfnamefont {L.}~\bibnamefont {Zhang}}, \ and\
  \bibinfo {author} {\bibfnamefont {C.}~\bibnamefont {Zhou}},\ }\href {\doibase
  10.1021/ar300203n} {\bibfield  {journal} {\bibinfo  {journal} {Acc. Chem.
  Res.}\ ,\ } (\bibinfo {year} {2013})}\BibitemShut {NoStop}%
\bibitem [{\citenamefont {Li}\ \emph {et~al.}(2009)\citenamefont {Li},
  \citenamefont {Cai}, \citenamefont {An}, \citenamefont {Kim}, \citenamefont
  {Nah}, \citenamefont {Yang}, \citenamefont {Piner}, \citenamefont
  {Velamakanni}, \citenamefont {Jung}, \citenamefont {Tutuc}, \citenamefont
  {Banerjee}, \citenamefont {Colombo},\ and\ \citenamefont {Ruoff}}]{Li2009c}%
  \BibitemOpen
  \bibfield  {author} {\bibinfo {author} {\bibfnamefont {X.}~\bibnamefont
  {Li}}, \bibinfo {author} {\bibfnamefont {W.}~\bibnamefont {Cai}}, \bibinfo
  {author} {\bibfnamefont {J.}~\bibnamefont {An}}, \bibinfo {author}
  {\bibfnamefont {S.}~\bibnamefont {Kim}}, \bibinfo {author} {\bibfnamefont
  {J.}~\bibnamefont {Nah}}, \bibinfo {author} {\bibfnamefont {D.}~\bibnamefont
  {Yang}}, \bibinfo {author} {\bibfnamefont {R.}~\bibnamefont {Piner}},
  \bibinfo {author} {\bibfnamefont {A.}~\bibnamefont {Velamakanni}}, \bibinfo
  {author} {\bibfnamefont {I.}~\bibnamefont {Jung}}, \bibinfo {author}
  {\bibfnamefont {E.}~\bibnamefont {Tutuc}}, \bibinfo {author} {\bibfnamefont
  {S.~K.}\ \bibnamefont {Banerjee}}, \bibinfo {author} {\bibfnamefont
  {L.}~\bibnamefont {Colombo}}, \ and\ \bibinfo {author} {\bibfnamefont
  {R.~S.}\ \bibnamefont {Ruoff}},\ }\href {\doibase 10.1126/science.1171245}
  {\bibfield  {journal} {\bibinfo  {journal} {Science}\ }\textbf {\bibinfo
  {volume} {324}},\ \bibinfo {pages} {1312} (\bibinfo {year}
  {2009})}\BibitemShut {NoStop}%
\bibitem [{\citenamefont {Bae}\ \emph {et~al.}(2010)\citenamefont {Bae},
  \citenamefont {Kim}, \citenamefont {Lee}, \citenamefont {Xu}, \citenamefont
  {Park}, \citenamefont {Zheng}, \citenamefont {Balakrishnan}, \citenamefont
  {Lei}, \citenamefont {Ri~Kim}, \citenamefont {Song}, \citenamefont {Kim},
  \citenamefont {Kim}, \citenamefont {Ozyilmaz}, \citenamefont {Ahn},
  \citenamefont {Hong},\ and\ \citenamefont {Iijima}}]{Bae2010}%
  \BibitemOpen
  \bibfield  {author} {\bibinfo {author} {\bibfnamefont {S.}~\bibnamefont
  {Bae}}, \bibinfo {author} {\bibfnamefont {H.}~\bibnamefont {Kim}}, \bibinfo
  {author} {\bibfnamefont {Y.}~\bibnamefont {Lee}}, \bibinfo {author}
  {\bibfnamefont {X.}~\bibnamefont {Xu}}, \bibinfo {author} {\bibfnamefont
  {J.-S.}\ \bibnamefont {Park}}, \bibinfo {author} {\bibfnamefont
  {Y.}~\bibnamefont {Zheng}}, \bibinfo {author} {\bibfnamefont
  {J.}~\bibnamefont {Balakrishnan}}, \bibinfo {author} {\bibfnamefont
  {T.}~\bibnamefont {Lei}}, \bibinfo {author} {\bibfnamefont {H.}~\bibnamefont
  {Ri~Kim}}, \bibinfo {author} {\bibfnamefont {Y.~I.}\ \bibnamefont {Song}},
  \bibinfo {author} {\bibfnamefont {Y.-J.}\ \bibnamefont {Kim}}, \bibinfo
  {author} {\bibfnamefont {K.~S.}\ \bibnamefont {Kim}}, \bibinfo {author}
  {\bibfnamefont {B.}~\bibnamefont {Ozyilmaz}}, \bibinfo {author}
  {\bibfnamefont {J.-H.}\ \bibnamefont {Ahn}}, \bibinfo {author} {\bibfnamefont
  {B.~H.}\ \bibnamefont {Hong}}, \ and\ \bibinfo {author} {\bibfnamefont
  {S.}~\bibnamefont {Iijima}},\ }\href {\doibase 10.1038/nnano.2010.132}
  {\bibfield  {journal} {\bibinfo  {journal} {Nat. Nanotechnol.}\ }\textbf
  {\bibinfo {volume} {5}},\ \bibinfo {pages} {574} (\bibinfo {year}
  {2010})}\BibitemShut {NoStop}%
\bibitem [{\citenamefont {Kobayashi}\ \emph {et~al.}(2013)\citenamefont
  {Kobayashi}, \citenamefont {Bando}, \citenamefont {Kimura}, \citenamefont
  {Shimizu}, \citenamefont {Kadono}, \citenamefont {Umezu}, \citenamefont
  {Miyahara}, \citenamefont {Hayazaki}, \citenamefont {Nagai}, \citenamefont
  {Mizuguchi}, \citenamefont {Murakami},\ and\ \citenamefont
  {Hobara}}]{Kobayashi2013}%
  \BibitemOpen
  \bibfield  {author} {\bibinfo {author} {\bibfnamefont {T.}~\bibnamefont
  {Kobayashi}}, \bibinfo {author} {\bibfnamefont {M.}~\bibnamefont {Bando}},
  \bibinfo {author} {\bibfnamefont {N.}~\bibnamefont {Kimura}}, \bibinfo
  {author} {\bibfnamefont {K.}~\bibnamefont {Shimizu}}, \bibinfo {author}
  {\bibfnamefont {K.}~\bibnamefont {Kadono}}, \bibinfo {author} {\bibfnamefont
  {N.}~\bibnamefont {Umezu}}, \bibinfo {author} {\bibfnamefont
  {K.}~\bibnamefont {Miyahara}}, \bibinfo {author} {\bibfnamefont
  {S.}~\bibnamefont {Hayazaki}}, \bibinfo {author} {\bibfnamefont
  {S.}~\bibnamefont {Nagai}}, \bibinfo {author} {\bibfnamefont
  {Y.}~\bibnamefont {Mizuguchi}}, \bibinfo {author} {\bibfnamefont
  {Y.}~\bibnamefont {Murakami}}, \ and\ \bibinfo {author} {\bibfnamefont
  {D.}~\bibnamefont {Hobara}},\ }\href {\doibase 10.1063/1.4776707} {\bibfield
  {journal} {\bibinfo  {journal} {Appl. Phys. Lett.}\ }\textbf {\bibinfo
  {volume} {102}},\ \bibinfo {pages} {023112} (\bibinfo {year}
  {2013})}\BibitemShut {NoStop}%
\bibitem [{\citenamefont {Li}\ \emph {et~al.}(2011)\citenamefont {Li},
  \citenamefont {Magnuson}, \citenamefont {Venugopal}, \citenamefont {Tromp},
  \citenamefont {Hannon}, \citenamefont {Vogel}, \citenamefont {Colombo},\ and\
  \citenamefont {Ruoff}}]{Li2011}%
  \BibitemOpen
  \bibfield  {author} {\bibinfo {author} {\bibfnamefont {X.}~\bibnamefont
  {Li}}, \bibinfo {author} {\bibfnamefont {C.~W.}\ \bibnamefont {Magnuson}},
  \bibinfo {author} {\bibfnamefont {A.}~\bibnamefont {Venugopal}}, \bibinfo
  {author} {\bibfnamefont {R.~M.}\ \bibnamefont {Tromp}}, \bibinfo {author}
  {\bibfnamefont {J.~B.}\ \bibnamefont {Hannon}}, \bibinfo {author}
  {\bibfnamefont {E.~M.}\ \bibnamefont {Vogel}}, \bibinfo {author}
  {\bibfnamefont {L.}~\bibnamefont {Colombo}}, \ and\ \bibinfo {author}
  {\bibfnamefont {R.~S.}\ \bibnamefont {Ruoff}},\ }\href {\doibase
  10.1021/ja109793s} {\bibfield  {journal} {\bibinfo  {journal} {J. Am. Chem.
  Soc.}\ }\textbf {\bibinfo {volume} {133}},\ \bibinfo {pages} {2816} (\bibinfo
  {year} {2011})}\BibitemShut {NoStop}%
\bibitem [{\citenamefont {Yan}\ \emph {et~al.}(2012)\citenamefont {Yan},
  \citenamefont {Lin}, \citenamefont {Peng}, \citenamefont {Sun}, \citenamefont
  {Zhu}, \citenamefont {Li}, \citenamefont {Xiang}, \citenamefont {Samuel},
  \citenamefont {Kittrell},\ and\ \citenamefont {Tour}}]{Yan2012}%
  \BibitemOpen
  \bibfield  {author} {\bibinfo {author} {\bibfnamefont {Z.}~\bibnamefont
  {Yan}}, \bibinfo {author} {\bibfnamefont {J.}~\bibnamefont {Lin}}, \bibinfo
  {author} {\bibfnamefont {Z.}~\bibnamefont {Peng}}, \bibinfo {author}
  {\bibfnamefont {Z.}~\bibnamefont {Sun}}, \bibinfo {author} {\bibfnamefont
  {Y.}~\bibnamefont {Zhu}}, \bibinfo {author} {\bibfnamefont {L.}~\bibnamefont
  {Li}}, \bibinfo {author} {\bibfnamefont {C.}~\bibnamefont {Xiang}}, \bibinfo
  {author} {\bibfnamefont {E.~L.}\ \bibnamefont {Samuel}}, \bibinfo {author}
  {\bibfnamefont {C.}~\bibnamefont {Kittrell}}, \ and\ \bibinfo {author}
  {\bibfnamefont {J.~M.}\ \bibnamefont {Tour}},\ }\bibfield  {booktitle} {\emph
  {\bibinfo {booktitle} {ACS Nano}},\ }\href {\doibase 10.1021/nn303352k}
  {\bibfield  {journal} {\bibinfo  {journal} {ACS Nano}\ }\textbf {\bibinfo
  {volume} {6}},\ \bibinfo {pages} {9110} (\bibinfo {year} {2012})}\BibitemShut
  {NoStop}%
\bibitem [{\citenamefont {Wu}\ \emph {et~al.}(2012)\citenamefont {Wu},
  \citenamefont {Fan}, \citenamefont {Speller}, \citenamefont {Creeth},
  \citenamefont {Sadowski}, \citenamefont {He}, \citenamefont {Robertson},
  \citenamefont {Allen},\ and\ \citenamefont {Warner}}]{Wu2012}%
  \BibitemOpen
  \bibfield  {author} {\bibinfo {author} {\bibfnamefont {Y.~A.}\ \bibnamefont
  {Wu}}, \bibinfo {author} {\bibfnamefont {Y.}~\bibnamefont {Fan}}, \bibinfo
  {author} {\bibfnamefont {S.}~\bibnamefont {Speller}}, \bibinfo {author}
  {\bibfnamefont {G.~L.}\ \bibnamefont {Creeth}}, \bibinfo {author}
  {\bibfnamefont {J.~T.}\ \bibnamefont {Sadowski}}, \bibinfo {author}
  {\bibfnamefont {K.}~\bibnamefont {He}}, \bibinfo {author} {\bibfnamefont
  {A.~W.}\ \bibnamefont {Robertson}}, \bibinfo {author} {\bibfnamefont {C.~S.}\
  \bibnamefont {Allen}}, \ and\ \bibinfo {author} {\bibfnamefont {J.~H.}\
  \bibnamefont {Warner}},\ }\bibfield  {booktitle} {\emph {\bibinfo {booktitle}
  {ACS Nano}},\ }\href {\doibase 10.1021/nn3016629} {\bibfield  {journal}
  {\bibinfo  {journal} {ACS Nano}\ }\textbf {\bibinfo {volume} {6}},\ \bibinfo
  {pages} {5010} (\bibinfo {year} {2012})}\BibitemShut {NoStop}%
\bibitem [{\citenamefont {Yu}\ \emph {et~al.}(2011)\citenamefont {Yu},
  \citenamefont {Jauregui}, \citenamefont {Wu}, \citenamefont {Colby},
  \citenamefont {Tian}, \citenamefont {Su}, \citenamefont {Cao}, \citenamefont
  {Liu}, \citenamefont {Pandey}, \citenamefont {Wei}, \citenamefont {Chung},
  \citenamefont {Peng}, \citenamefont {Guisinger}, \citenamefont {Stach},
  \citenamefont {Bao}, \citenamefont {Pei},\ and\ \citenamefont
  {Chen}}]{Yu2011}%
  \BibitemOpen
  \bibfield  {author} {\bibinfo {author} {\bibfnamefont {Q.}~\bibnamefont
  {Yu}}, \bibinfo {author} {\bibfnamefont {L.~A.}\ \bibnamefont {Jauregui}},
  \bibinfo {author} {\bibfnamefont {W.}~\bibnamefont {Wu}}, \bibinfo {author}
  {\bibfnamefont {R.}~\bibnamefont {Colby}}, \bibinfo {author} {\bibfnamefont
  {J.}~\bibnamefont {Tian}}, \bibinfo {author} {\bibfnamefont {Z.}~\bibnamefont
  {Su}}, \bibinfo {author} {\bibfnamefont {H.}~\bibnamefont {Cao}}, \bibinfo
  {author} {\bibfnamefont {Z.}~\bibnamefont {Liu}}, \bibinfo {author}
  {\bibfnamefont {D.}~\bibnamefont {Pandey}}, \bibinfo {author} {\bibfnamefont
  {D.}~\bibnamefont {Wei}}, \bibinfo {author} {\bibfnamefont {T.~F.}\
  \bibnamefont {Chung}}, \bibinfo {author} {\bibfnamefont {P.}~\bibnamefont
  {Peng}}, \bibinfo {author} {\bibfnamefont {N.~P.}\ \bibnamefont {Guisinger}},
  \bibinfo {author} {\bibfnamefont {E.~A.}\ \bibnamefont {Stach}}, \bibinfo
  {author} {\bibfnamefont {J.}~\bibnamefont {Bao}}, \bibinfo {author}
  {\bibfnamefont {S.-S.}\ \bibnamefont {Pei}}, \ and\ \bibinfo {author}
  {\bibfnamefont {Y.~P.}\ \bibnamefont {Chen}},\ }\href {\doibase
  10.1038/nmat3010} {\bibfield  {journal} {\bibinfo  {journal} {Nat. Mater.}\
  }\textbf {\bibinfo {volume} {10}},\ \bibinfo {pages} {443} (\bibinfo {year}
  {2011})}\BibitemShut {NoStop}%
\bibitem [{\citenamefont {Koepke}\ \emph {et~al.}(2012)\citenamefont {Koepke},
  \citenamefont {Wood}, \citenamefont {Estrada}, \citenamefont {Ong},
  \citenamefont {He}, \citenamefont {Pop},\ and\ \citenamefont
  {Lyding}}]{Koepke2012}%
  \BibitemOpen
  \bibfield  {author} {\bibinfo {author} {\bibfnamefont {J.~C.}\ \bibnamefont
  {Koepke}}, \bibinfo {author} {\bibfnamefont {J.~D.}\ \bibnamefont {Wood}},
  \bibinfo {author} {\bibfnamefont {D.}~\bibnamefont {Estrada}}, \bibinfo
  {author} {\bibfnamefont {Z.-Y.}\ \bibnamefont {Ong}}, \bibinfo {author}
  {\bibfnamefont {K.~T.}\ \bibnamefont {He}}, \bibinfo {author} {\bibfnamefont
  {E.}~\bibnamefont {Pop}}, \ and\ \bibinfo {author} {\bibfnamefont {J.~W.}\
  \bibnamefont {Lyding}},\ }\bibfield  {booktitle} {\emph {\bibinfo {booktitle}
  {ACS Nano}},\ }\href {\doibase 10.1021/nn302064p} {\bibfield  {journal}
  {\bibinfo  {journal} {ACS Nano}\ }\textbf {\bibinfo {volume} {7}},\ \bibinfo
  {pages} {75} (\bibinfo {year} {2012})}\BibitemShut {NoStop}%
\bibitem [{\citenamefont {Li}\ \emph {et~al.}(2010)\citenamefont {Li},
  \citenamefont {Magnuson}, \citenamefont {Venugopal}, \citenamefont {An},
  \citenamefont {Suk}, \citenamefont {Han}, \citenamefont {Borysiak},
  \citenamefont {Cai}, \citenamefont {Velamakanni}, \citenamefont {Zhu},
  \citenamefont {Fu}, \citenamefont {Vogel}, \citenamefont {Voelkl},
  \citenamefont {Colombo},\ and\ \citenamefont {Ruoff}}]{Li2010}%
  \BibitemOpen
  \bibfield  {author} {\bibinfo {author} {\bibfnamefont {X.}~\bibnamefont
  {Li}}, \bibinfo {author} {\bibfnamefont {C.~W.}\ \bibnamefont {Magnuson}},
  \bibinfo {author} {\bibfnamefont {A.}~\bibnamefont {Venugopal}}, \bibinfo
  {author} {\bibfnamefont {J.}~\bibnamefont {An}}, \bibinfo {author}
  {\bibfnamefont {J.~W.}\ \bibnamefont {Suk}}, \bibinfo {author} {\bibfnamefont
  {B.}~\bibnamefont {Han}}, \bibinfo {author} {\bibfnamefont {M.}~\bibnamefont
  {Borysiak}}, \bibinfo {author} {\bibfnamefont {W.}~\bibnamefont {Cai}},
  \bibinfo {author} {\bibfnamefont {A.}~\bibnamefont {Velamakanni}}, \bibinfo
  {author} {\bibfnamefont {Y.}~\bibnamefont {Zhu}}, \bibinfo {author}
  {\bibfnamefont {L.}~\bibnamefont {Fu}}, \bibinfo {author} {\bibfnamefont
  {E.~M.}\ \bibnamefont {Vogel}}, \bibinfo {author} {\bibfnamefont
  {E.}~\bibnamefont {Voelkl}}, \bibinfo {author} {\bibfnamefont
  {L.}~\bibnamefont {Colombo}}, \ and\ \bibinfo {author} {\bibfnamefont
  {R.~S.}\ \bibnamefont {Ruoff}},\ }\bibfield  {booktitle} {\emph {\bibinfo
  {booktitle} {Nano Letters}},\ }\href {\doibase 10.1021/nl101629g} {\bibfield
  {journal} {\bibinfo  {journal} {Nano Lett.}\ }\textbf {\bibinfo {volume}
  {10}},\ \bibinfo {pages} {4328} (\bibinfo {year} {2010})}\BibitemShut
  {NoStop}%
\bibitem [{\citenamefont {Yazyev}\ and\ \citenamefont
  {Louie}(2010)}]{Yazyev2010}%
  \BibitemOpen
  \bibfield  {author} {\bibinfo {author} {\bibfnamefont {O.~V.}\ \bibnamefont
  {Yazyev}}\ and\ \bibinfo {author} {\bibfnamefont {S.~G.}\ \bibnamefont
  {Louie}},\ }\href {\doibase 10.1038/nmat2830} {\bibfield  {journal} {\bibinfo
   {journal} {Nat. Mater.}\ }\textbf {\bibinfo {volume} {9}},\ \bibinfo {pages}
  {806} (\bibinfo {year} {2010})}\BibitemShut {NoStop}%
\bibitem [{\citenamefont {Petrone}\ \emph {et~al.}(2012)\citenamefont
  {Petrone}, \citenamefont {Dean}, \citenamefont {Meric}, \citenamefont
  {van~der Zande}, \citenamefont {Huang}, \citenamefont {Wang}, \citenamefont
  {Muller}, \citenamefont {Shepard},\ and\ \citenamefont {Hone}}]{Petrone2012}%
  \BibitemOpen
  \bibfield  {author} {\bibinfo {author} {\bibfnamefont {N.}~\bibnamefont
  {Petrone}}, \bibinfo {author} {\bibfnamefont {C.~R.}\ \bibnamefont {Dean}},
  \bibinfo {author} {\bibfnamefont {I.}~\bibnamefont {Meric}}, \bibinfo
  {author} {\bibfnamefont {A.~M.}\ \bibnamefont {van~der Zande}}, \bibinfo
  {author} {\bibfnamefont {P.~Y.}\ \bibnamefont {Huang}}, \bibinfo {author}
  {\bibfnamefont {L.}~\bibnamefont {Wang}}, \bibinfo {author} {\bibfnamefont
  {D.}~\bibnamefont {Muller}}, \bibinfo {author} {\bibfnamefont {K.~L.}\
  \bibnamefont {Shepard}}, \ and\ \bibinfo {author} {\bibfnamefont
  {J.}~\bibnamefont {Hone}},\ }\bibfield  {booktitle} {\emph {\bibinfo
  {booktitle} {Nano Letters}},\ }\href {\doibase 10.1021/nl204481s} {\bibfield
  {journal} {\bibinfo  {journal} {Nano Lett.}\ }\textbf {\bibinfo {volume}
  {12}},\ \bibinfo {pages} {2751} (\bibinfo {year} {2012})}\BibitemShut
  {NoStop}%
\bibitem [{\citenamefont {Gannett}\ \emph {et~al.}(2011)\citenamefont
  {Gannett}, \citenamefont {Regan}, \citenamefont {Watanabe}, \citenamefont
  {Taniguchi}, \citenamefont {Crommie},\ and\ \citenamefont
  {Zettl}}]{Gannett2011}%
  \BibitemOpen
  \bibfield  {author} {\bibinfo {author} {\bibfnamefont {W.}~\bibnamefont
  {Gannett}}, \bibinfo {author} {\bibfnamefont {W.}~\bibnamefont {Regan}},
  \bibinfo {author} {\bibfnamefont {K.}~\bibnamefont {Watanabe}}, \bibinfo
  {author} {\bibfnamefont {T.}~\bibnamefont {Taniguchi}}, \bibinfo {author}
  {\bibfnamefont {M.~F.}\ \bibnamefont {Crommie}}, \ and\ \bibinfo {author}
  {\bibfnamefont {A.}~\bibnamefont {Zettl}},\ }\href {\doibase
  10.1063/1.3599708} {\bibfield  {journal} {\bibinfo  {journal} {Appl. Phys.
  Lett.}\ }\textbf {\bibinfo {volume} {98}},\ \bibinfo {pages} {242105}
  (\bibinfo {year} {2011})}\BibitemShut {NoStop}%
\bibitem [{\citenamefont {Zomer}\ \emph {et~al.}(2011)\citenamefont {Zomer},
  \citenamefont {Dash}, \citenamefont {Tombros},\ and\ \citenamefont {van
  Wees}}]{Zomer2011}%
  \BibitemOpen
  \bibfield  {author} {\bibinfo {author} {\bibfnamefont {P.~J.}\ \bibnamefont
  {Zomer}}, \bibinfo {author} {\bibfnamefont {S.~P.}\ \bibnamefont {Dash}},
  \bibinfo {author} {\bibfnamefont {N.}~\bibnamefont {Tombros}}, \ and\
  \bibinfo {author} {\bibfnamefont {B.~J.}\ \bibnamefont {van Wees}},\ }\href
  {\doibase 10.1063/1.3665405} {\bibfield  {journal} {\bibinfo  {journal}
  {Appl. Phys. Lett.}\ }\textbf {\bibinfo {volume} {99}},\ \bibinfo {pages}
  {232104} (\bibinfo {year} {2011})}\BibitemShut {NoStop}%
\bibitem [{\citenamefont {Cheianov}, \citenamefont {Fal'ko},\ and\
  \citenamefont {Altshuler}(2007)}]{Cheianov2007}%
  \BibitemOpen
  \bibfield  {author} {\bibinfo {author} {\bibfnamefont {V.~V.}\ \bibnamefont
  {Cheianov}}, \bibinfo {author} {\bibfnamefont {V.}~\bibnamefont {Fal'ko}}, \
  and\ \bibinfo {author} {\bibfnamefont {B.~L.}\ \bibnamefont {Altshuler}},\
  }\href {\doibase 10.1126/science.1138020} {\bibfield  {journal} {\bibinfo
  {journal} {Science}\ }\textbf {\bibinfo {volume} {315}},\ \bibinfo {pages}
  {1252} (\bibinfo {year} {2007})}\BibitemShut {NoStop}%
\bibitem [{\citenamefont {Katsnelson}, \citenamefont {Novoselov},\ and\
  \citenamefont {Geim}(2006)}]{Katsnelson2006}%
  \BibitemOpen
  \bibfield  {author} {\bibinfo {author} {\bibfnamefont {M.~I.}\ \bibnamefont
  {Katsnelson}}, \bibinfo {author} {\bibfnamefont {K.~S.}\ \bibnamefont
  {Novoselov}}, \ and\ \bibinfo {author} {\bibfnamefont {A.~K.}\ \bibnamefont
  {Geim}},\ }\href {\doibase 10.1038/nphys384} {\bibfield  {journal} {\bibinfo
  {journal} {Nat. Phys.}\ }\textbf {\bibinfo {volume} {2}},\ \bibinfo {pages}
  {620} (\bibinfo {year} {2006})}\BibitemShut {NoStop}%
\bibitem [{\citenamefont {Beenakker}(2006)}]{Beenakker2006}%
  \BibitemOpen
  \bibfield  {author} {\bibinfo {author} {\bibfnamefont {C.~W.~J.}\
  \bibnamefont {Beenakker}},\ }\href {\doibase 10.1103/PhysRevLett.97.067007}
  {\bibfield  {journal} {\bibinfo  {journal} {Phys. Rev. Lett.}\ }\textbf
  {\bibinfo {volume} {97}},\ \bibinfo {pages} {067007} (\bibinfo {year}
  {2006})}\BibitemShut {NoStop}%
\bibitem [{\citenamefont {Hirayama}\ \emph {et~al.}(1991)\citenamefont
  {Hirayama}, \citenamefont {Saku}, \citenamefont {Tarucha},\ and\
  \citenamefont {Horikoshi}}]{Hirayama1991}%
  \BibitemOpen
  \bibfield  {author} {\bibinfo {author} {\bibfnamefont {Y.}~\bibnamefont
  {Hirayama}}, \bibinfo {author} {\bibfnamefont {T.}~\bibnamefont {Saku}},
  \bibinfo {author} {\bibfnamefont {S.}~\bibnamefont {Tarucha}}, \ and\
  \bibinfo {author} {\bibfnamefont {Y.}~\bibnamefont {Horikoshi}},\ }\href
  {http://dx.doi.org/10.1063/1.104803} {\bibfield  {journal} {\bibinfo
  {journal} {Appl. Phys. Lett.}\ }\textbf {\bibinfo {volume} {58}},\ \bibinfo
  {pages} {2672} (\bibinfo {year} {1991})}\BibitemShut {NoStop}%
\bibitem [{\citenamefont {Mayorov}\ \emph {et~al.}(2011)\citenamefont
  {Mayorov}, \citenamefont {Gorbachev}, \citenamefont {Morozov}, \citenamefont
  {Britnell}, \citenamefont {Jalil}, \citenamefont {Ponomarenko}, \citenamefont
  {Blake}, \citenamefont {Novoselov}, \citenamefont {Watanabe}, \citenamefont
  {Taniguchi},\ and\ \citenamefont {Geim}}]{Mayorov2011}%
  \BibitemOpen
  \bibfield  {author} {\bibinfo {author} {\bibfnamefont {A.~S.}\ \bibnamefont
  {Mayorov}}, \bibinfo {author} {\bibfnamefont {R.~V.}\ \bibnamefont
  {Gorbachev}}, \bibinfo {author} {\bibfnamefont {S.~V.}\ \bibnamefont
  {Morozov}}, \bibinfo {author} {\bibfnamefont {L.}~\bibnamefont {Britnell}},
  \bibinfo {author} {\bibfnamefont {R.}~\bibnamefont {Jalil}}, \bibinfo
  {author} {\bibfnamefont {L.~A.}\ \bibnamefont {Ponomarenko}}, \bibinfo
  {author} {\bibfnamefont {P.}~\bibnamefont {Blake}}, \bibinfo {author}
  {\bibfnamefont {K.~S.}\ \bibnamefont {Novoselov}}, \bibinfo {author}
  {\bibfnamefont {K.}~\bibnamefont {Watanabe}}, \bibinfo {author}
  {\bibfnamefont {T.}~\bibnamefont {Taniguchi}}, \ and\ \bibinfo {author}
  {\bibfnamefont {A.~K.}\ \bibnamefont {Geim}},\ }\href {\doibase
  10.1021/nl200758b} {\bibfield  {journal} {\bibinfo  {journal} {Nano Lett.}\
  }\textbf {\bibinfo {volume} {11}},\ \bibinfo {pages} {2396} (\bibinfo {year}
  {2011})}\BibitemShut {NoStop}%
\bibitem [{\citenamefont {van Houten}\ \emph {et~al.}(1989)\citenamefont {van
  Houten}, \citenamefont {Beenakker}, \citenamefont {Williamson}, \citenamefont
  {Broekaart}, \citenamefont {van Loosdrecht}, \citenamefont {van Wees},
  \citenamefont {Mooij}, \citenamefont {Foxon},\ and\ \citenamefont
  {Harris}}]{Houten1989}%
  \BibitemOpen
  \bibfield  {author} {\bibinfo {author} {\bibfnamefont {H.}~\bibnamefont {van
  Houten}}, \bibinfo {author} {\bibfnamefont {C.~W.~J.}\ \bibnamefont
  {Beenakker}}, \bibinfo {author} {\bibfnamefont {J.~G.}\ \bibnamefont
  {Williamson}}, \bibinfo {author} {\bibfnamefont {M.~E.~I.}\ \bibnamefont
  {Broekaart}}, \bibinfo {author} {\bibfnamefont {P.~H.~M.}\ \bibnamefont {van
  Loosdrecht}}, \bibinfo {author} {\bibfnamefont {B.~J.}\ \bibnamefont {van
  Wees}}, \bibinfo {author} {\bibfnamefont {J.~E.}\ \bibnamefont {Mooij}},
  \bibinfo {author} {\bibfnamefont {C.~T.}\ \bibnamefont {Foxon}}, \ and\
  \bibinfo {author} {\bibfnamefont {J.~J.}\ \bibnamefont {Harris}},\ }\href
  {\doibase 10.1103/PhysRevB.39.8556} {\bibfield  {journal} {\bibinfo
  {journal} {Phys. Rev. B}\ }\textbf {\bibinfo {volume} {39}},\ \bibinfo
  {pages} {8556} (\bibinfo {year} {1989})}\BibitemShut {NoStop}%
\bibitem [{\citenamefont {Taychatanapat}\ \emph {et~al.}(2013)\citenamefont
  {Taychatanapat}, \citenamefont {Watanabe}, \citenamefont {Taniguchi},\ and\
  \citenamefont {Jarillo-Herrero}}]{Taychatanapat2013}%
  \BibitemOpen
  \bibfield  {author} {\bibinfo {author} {\bibfnamefont {T.}~\bibnamefont
  {Taychatanapat}}, \bibinfo {author} {\bibfnamefont {K.}~\bibnamefont
  {Watanabe}}, \bibinfo {author} {\bibfnamefont {T.}~\bibnamefont {Taniguchi}},
  \ and\ \bibinfo {author} {\bibfnamefont {P.}~\bibnamefont
  {Jarillo-Herrero}},\ }\href {\doibase 10.1038/nphys2549} {\bibfield
  {journal} {\bibinfo  {journal} {Nat. Phys.}\ }\textbf {\bibinfo {volume}
  {9}},\ \bibinfo {pages} {225} (\bibinfo {year} {2013})}\BibitemShut {NoStop}%
\bibitem [{\citenamefont {Ferrari}\ \emph {et~al.}(2006)\citenamefont
  {Ferrari}, \citenamefont {Meyer}, \citenamefont {Scardaci}, \citenamefont
  {Casiraghi}, \citenamefont {Lazzeri}, \citenamefont {Mauri}, \citenamefont
  {Piscanec}, \citenamefont {Jiang}, \citenamefont {Novoselov}, \citenamefont
  {Roth},\ and\ \citenamefont {Geim}}]{Ferrari2006}%
  \BibitemOpen
  \bibfield  {author} {\bibinfo {author} {\bibfnamefont {A.~C.}\ \bibnamefont
  {Ferrari}}, \bibinfo {author} {\bibfnamefont {J.~C.}\ \bibnamefont {Meyer}},
  \bibinfo {author} {\bibfnamefont {V.}~\bibnamefont {Scardaci}}, \bibinfo
  {author} {\bibfnamefont {C.}~\bibnamefont {Casiraghi}}, \bibinfo {author}
  {\bibfnamefont {M.}~\bibnamefont {Lazzeri}}, \bibinfo {author} {\bibfnamefont
  {F.}~\bibnamefont {Mauri}}, \bibinfo {author} {\bibfnamefont
  {S.}~\bibnamefont {Piscanec}}, \bibinfo {author} {\bibfnamefont
  {D.}~\bibnamefont {Jiang}}, \bibinfo {author} {\bibfnamefont {K.~S.}\
  \bibnamefont {Novoselov}}, \bibinfo {author} {\bibfnamefont {S.}~\bibnamefont
  {Roth}}, \ and\ \bibinfo {author} {\bibfnamefont {A.~K.}\ \bibnamefont
  {Geim}},\ }\href {\doibase 10.1103/PhysRevLett.97.187401} {\bibfield
  {journal} {\bibinfo  {journal} {Phys. Rev. Lett.}\ }\textbf {\bibinfo
  {volume} {97}},\ \bibinfo {pages} {187401} (\bibinfo {year}
  {2006})}\BibitemShut {NoStop}%
\bibitem [{\citenamefont {Meyer}\ \emph {et~al.}(2007)\citenamefont {Meyer},
  \citenamefont {Geim}, \citenamefont {Katsnelson}, \citenamefont {Novoselov},
  \citenamefont {Booth},\ and\ \citenamefont {Roth}}]{Meyer2007}%
  \BibitemOpen
  \bibfield  {author} {\bibinfo {author} {\bibfnamefont {J.~C.}\ \bibnamefont
  {Meyer}}, \bibinfo {author} {\bibfnamefont {A.~K.}\ \bibnamefont {Geim}},
  \bibinfo {author} {\bibfnamefont {M.~I.}\ \bibnamefont {Katsnelson}},
  \bibinfo {author} {\bibfnamefont {K.~S.}\ \bibnamefont {Novoselov}}, \bibinfo
  {author} {\bibfnamefont {T.~J.}\ \bibnamefont {Booth}}, \ and\ \bibinfo
  {author} {\bibfnamefont {S.}~\bibnamefont {Roth}},\ }\href {\doibase
  10.1038/nature05545} {\bibfield  {journal} {\bibinfo  {journal} {Nature
  (London)}\ }\textbf {\bibinfo {volume} {446}},\ \bibinfo {pages} {60}
  (\bibinfo {year} {2007})}\BibitemShut {NoStop}%
\bibitem [{Sup()}]{Supplement}%
  \BibitemOpen
  \href@noop {} {\bibinfo  {journal} {See supplementary material at [URL will
  be inserted by AIP] for further details and analysis.}\ }\BibitemShut
  {NoStop}%
\bibitem [{\citenamefont {Adam}\ \emph {et~al.}(2007)\citenamefont {Adam},
  \citenamefont {Hwang}, \citenamefont {Galitski},\ and\ \citenamefont
  {Das~Sarma}}]{Adam2007}%
  \BibitemOpen
\bibfield  {journal} {  }\bibfield  {author} {\bibinfo {author} {\bibfnamefont
  {S.}~\bibnamefont {Adam}}, \bibinfo {author} {\bibfnamefont {E.~H.}\
  \bibnamefont {Hwang}}, \bibinfo {author} {\bibfnamefont {V.~M.}\ \bibnamefont
  {Galitski}}, \ and\ \bibinfo {author} {\bibfnamefont {S.}~\bibnamefont
  {Das~Sarma}},\ }\href {\doibase 10.1073/pnas.0704772104} {\bibfield
  {journal} {\bibinfo  {journal} {P. Natl. A. Sci. USA}\ }\textbf {\bibinfo
  {volume} {104}},\ \bibinfo {pages} {18392} (\bibinfo {year}
  {2007})}\BibitemShut {NoStop}%
\bibitem [{\citenamefont {Hwang}, \citenamefont {Adam},\ and\ \citenamefont
  {Das~Sarma}(2007)}]{Hwang2007}%
  \BibitemOpen
  \bibfield  {author} {\bibinfo {author} {\bibfnamefont {E.~H.}\ \bibnamefont
  {Hwang}}, \bibinfo {author} {\bibfnamefont {S.}~\bibnamefont {Adam}}, \ and\
  \bibinfo {author} {\bibfnamefont {S.}~\bibnamefont {Das~Sarma}},\ }\href
  {\doibase 10.1103/PhysRevLett.98.186806} {\bibfield  {journal} {\bibinfo
  {journal} {Phys. Rev. Lett.}\ }\textbf {\bibinfo {volume} {98}},\ \bibinfo
  {pages} {186806} (\bibinfo {year} {2007})}\BibitemShut {NoStop}%
\end{thebibliography}
\end{document}